\documentclass[11pt,a4paper,twoside,groupcitations]{article}
\usepackage[T1]{fontenc}
\usepackage[ansinew]{inputenc}
\usepackage[english]{babel}
\usepackage{amsfonts}
\usepackage{amsmath}
\usepackage{bm}
\usepackage{array,upgreek,accents}
\usepackage{amsthm}
\usepackage{amssymb}
\usepackage{graphicx}
\usepackage{braket}
\usepackage{eucal}
\usepackage{verbatim}
\usepackage[table]{xcolor}
\usepackage{caption}
\usepackage{cite}
\usepackage{textcomp}
\usepackage{comment}
\usepackage{float}
\usepackage{multicol}
\usepackage{tikz}
\usetikzlibrary{positioning,arrows}
\usetikzlibrary{decorations.pathmorphing}
\usetikzlibrary{decorations.markings}
\usetikzlibrary{calc,decorations.markings}
\usetikzlibrary{arrows,shapes}
\usetikzlibrary{matrix,arrows}

\usepackage[colorlinks,hyperindex,unicode]{hyperref}
\definecolor{green1}{RGB}{0,128,0} 
\hypersetup{hidelinks,backref=true,pagebackref=true,hyperindex=true,colorlinks=true,breaklinks=true,urlcolor= blue}
\hypersetup{%
  colorlinks = true,
  linkcolor  = blue,
  citecolor = cyan,
}

\newcommand{\beq}{\begin{eqnarray}}
\newcommand{\eeq}{\end{eqnarray}}
\newcommand{\be}{\begin{eqnarray}}
\newcommand{\ee}{\end{eqnarray}}

\renewcommand{\d}{\mbox{${\rm d}$}} 
\newcommand{\lp}{\ell_{\rm p}}
\newcommand{\mpl}{m_{\rm p}}
\newcommand{\gn}{G_{\rm N}}

\title{\bf Quantum maximally symmetric spacetimes}
\author{P.~Meert$^{a}$\thanks{E-mail: pedro.meert@unesp.br},
$\ $
A.~Giusti$^{b}$\thanks{E-mail: A.Giusti@sussex.ac.uk},
$\ $
and
R.~Casadio$^{cd}$\thanks{E-mail: casadio@bo.infn.it}
\\
\\
$^a${\em Instituto de F\'isica Te\'orica, Unesp,}
\\
{\em S\~ao Paulo, 01140-070, Brazil}
\\
\\
$^b$ {\em Department of Physics and Astronomy}
\\
{\em University of Sussex, Brighton, BN1~9QH, United Kingdom}
\\
\\
$^c${\em Dipartimento di Fisica e Astronomia, Universit\`a di Bologna}
\\
{\em via Irnerio~46, 40126 Bologna, Italy}
\\
\\
$^d${\em I.N.F.N., Sezione di Bologna, I.S.~FLAG}
\\
{\em viale B.~Pichat~6/2, 40127 Bologna, Italy}
}
\begin{document}
\maketitle
\begin{abstract}
	We show that 4-dimensional maximally symmetric spacetimes can be obtained
	from a coherent state quantisation of gravity, always resulting in geometries
	that approach the Minkowski vacuum exponentially away from the radius of
	curvature.
	A possible connection with the central charge in the AdS/CFT correspondence
	is also noted.
\end{abstract}
\newpage
\section{Introduction}
\setcounter{equation}{0}
\label{Sect:Intro}
The simplest solutions of the Einstein equations are maximally symmetric spacetimes.
They arise once we assume the manifold admits all possible symmetries, hence
they admit $n\left( n-1 \right)/2$ Killing vectors if $n$ is the spacetime dimension~\cite{weinberg}.
This condition highly constrains the geometry.
For instance, in $n=4$, the Riemann tensor reads
\begin{equation}
	\label{eq:Riemann}
	R_{abcd}
	=k
	\left(g_{ac}\,g_{bd}-g_{ad}\,g_{bc}\right)
	\  ,
\end{equation}
where
\begin{equation}
	k=
	\frac{R}{12}
	=
	\frac{1}{L^{2}}
	\ ,
\end{equation}
and the scalar curvature $R$ is a constant ($L$ is the radius of curvature) which
completely determines the spacetime.
In fact, for the Riemann tensor~\eqref{eq:Riemann}, the vacuum Einstein equations in the
presence of the cosmological constant $\Lambda$ read
\begin{equation}
	\label{eq:Einstein_eq}
	R
	=
	4\,\Lambda
	=
	\frac{12}{L^{2}}
	\ ,
\end{equation}
and we have three distinct solutions given by $\Lambda=0$, $\Lambda>0$ and $\Lambda<0$,
corresponding to Minkowski, de~Sitter (dS) and anti-de~Sitter (AdS) spacetimes, respectively.
\par
The dS spacetime is especially important in cosmology, where it has inspired inflationary
models and is used to describe the late time acceleration of the Universe.
This is possible because the dS spacetime contains an ever expanding space-like section,
but for most of this work we are going to use coordinates adapted to the static patch with
the metric given by
\begin{equation}
	\d s^{2}
	=
	-
	\left(1-\frac{r^{2}}{L^{2}}\right)\d t^{2}
	+
	\left(1-\frac{r^{2}}{L^{2}}\right)^{-1}
	\d r^{2}
	+
	r^{2}\,\d\Omega^{2}
	\ .
	\label{gds}
\end{equation}
Such coordinates do not cover the spacetime entirely as they are limited by a cosmological
horizon located at $r=L=\sqrt{3/\Lambda}$.
\par
The AdS spacetime has attracted attention particularly since the AdS/CFT conjecture relates
gravitational physics in AdS backgrounds to conformal field theories without gravity on the border,
which allows for certain quantum gravity computations in toy models.
The AdS/CFT correspondence also has some applications to plasma and condensed matter
physics in more elaborate setups~\cite{Ammon:2015wua}.
The metric for the AdS spacetime can be written as
\begin{equation}
	\d s^{2}
	=
	-
	\left(1+\frac{r^{2}}{L^{2}}\right)\d t^{2}
	+
	\left(1+\frac{r^{2}}{L^{2}}\right)^{-1}
	\d r^{2}
	+
	r^{2}\,\d\Omega^{2}
	\ ,
	\label{gads}
\end{equation}
where the spacetime is now covered entirely, as $L = \sqrt{3/|\Lambda|}>0$ (with $ \Lambda<0$)
and no horizons are present.
\par
Among many attempts at quantising gravity, the corpuscular approach is a simple prescription where the
geometry of spacetime is seen as an emergent property of the more fundamental, collective behaviour, of
gravitons described by quantum field theory in flat Minkowski spacetime~\cite{Dvali:2011aa,Dvali:2013eja}.
That approach has inspired~\cite{Casadio:2016zpl} the coherent quantisation of gravity~\cite{Casadio:2021eio}
in which the spacetime manifold and geometry emerge as a ``creation'' process from a vacuum formally
represented by a Minkowski spacetime (of unspecified dimension) which gives rise to coherent
states~\cite{Muck:2013orm,Berezhiani:2021zst,Oriti:2021oux} that reproduce solutions of the classical
Einstein equations as closely as possible.
Black holes and their associated quantum corrected geometries have been analysed in this approach,
including a cosmological constant~\cite{Giusti:2021shf}, electric charge~\cite{Casadio:2022ndh} and
rotation~\cite{Casadio:2023iqt,Feng:2024bsx}, their horizon~\cite{Feng:2024nvv} and
thermodynamics~\cite{Casadio:2021eio,Feng:2024bsx,Casadio:2023pmh}, and potential observational
signatures~\cite{Urmanov:2024qai}.
\par
The only solutions of the Einstein equations with a cosmological constant analysed so far is the
Schwarzschild-dS spacetime, as a model for dark matter effects (see the discussion in Ref.~\cite{Giusti:2021shf},
which was later refined in~\cite{Nancy} applying a different regularisation procedure compared
to the one discussed in this work).
In light of the relevance of dS and AdS geometries in current research areas, we are going to
investigate the quantum corrected metrics of these maximally symmetric spacetimes, and
address relevant questions, in particular the role played by the cosmological constant in
quantum gravity.
\par
In Section~\ref{Sect:Qgeom}, we introduce the general formalism used to obtain the quantum-corrected geometry
by constructing coherent states.
We next specialise the construction of Section.~\ref{Sect:Qgeom} to maximally symmetric spacetimes
in Section~\ref{Sect:QAdS}, where we also investigate their properties;
We end with a summary and discussion of the main results in Section~\ref{Sect:concl}.
We use units such that the speed of light $ c = 1 $, Newton's constant $\gn={\lp}/{\mpl} $
and Planck's constants $ \hbar = \mpl\,\lp $, with $ \lp $ and $ \mpl $ the Planck mass and
length, respectively.
\section{Coherent states of the gravitational field}
\setcounter{equation}{0}
\label{Sect:Qgeom}
In the coherent state quantisation of gravity, the spacetime geometry should emerge as the
mean field of the metric tensor $\bm g=\bra{g} \hat {\bm g} \ket{g}$.
The state $\ket{g}$ is a coherent state obtained by turning on gravitational modes that
generate the spacetime manifold from a vacuum state $\ket{0}$ which can be formally
associated with a Minkowski geometry describing no physical events.
\par
For static and spherically symmetric geometries, whose components can be written as
\begin{equation}
	\label{g}
	-g_{tt}
	=
	g^{rr}
	=
	1+V(r)
	\ ,
\end{equation}
it is sufficient to consider a single gravitational mode represented by a massless
(canonically normalised) scalar field $\Phi $ satisfying the vacuum equation of motion
\begin{equation}
	\label{eq:KGfull}
	\Box\Phi
	=
	\left[
		-
		\frac{\partial^2}{\partial t^2}
		+
		\frac{1}{r^{2}}\,\frac{\partial}{\partial r}
		\left(r^{2}\,\frac{\partial}{\partial r}\right)
		+
		\frac{1}{r^{2}\sin\theta}\,\frac{\partial}{\partial \theta}
		\left(\sin\theta\,\frac{\partial}{\partial \theta}\right)
		+
		\frac{1}{r^{2}\sin^{2}\theta}
		\frac{\partial^2}{\partial \phi^2}
		\right]\Phi
	=
	0
	\ .
\end{equation}
Since the metric function $V=V(r)$ that we wish to reproduce has no angular dependence,
we can assume $\Phi=\Phi(t,r)$, which yields the normal modes
\begin{equation}
	u_{k}(r,t)
	=
	j_{0}(k\,r)\,e^{-\,i\,k\,t}
	\ ,
\end{equation}
where
\begin{equation}
	j_{0}(k\,r)
	=
	\frac{\sin\left(kr\right)}{kr}
\end{equation}
are spherical Bessel functions of order zero satisfying the orthogonality relation
\begin{equation}
	\label{eq:bessel_ortog}
	4\,\pi
	\int_{0}^{\infty}
	r^{2}\,\d r\,
	j_0(k\,r)\,j_0(p\,r)
	=
	\frac{2\,\pi^{2}}{k^{2}}\,
	\delta(k-p)
	\iff
	\int_{0}^{\infty}
	\frac{k^{2}\,\d k}{2\,\pi^{2}}
	j_{0}\left(k\,r\right)j_{0}\left(k\,s\right)
	=
	\frac{1}{4\,\pi\, r^{2}}\,\delta\left(r-s\right)
	\ .
\end{equation}
Promoting the scalar field to an operator leads to
\begin{equation}
	\hat{\Phi}(r,t)
	=
	\int_0^\infty
	\frac{k^{2}\,\d k}{2\,\pi^{2}}\,
	\sqrt{\frac{\hbar}{2\,k}}
	\left(u_{k}\,\hat{a}_{k}
	+u_{k}^{*}\,\hat{a}_{k}^{\dagger}
	\right)
\end{equation}
and the conjugate momentum $\hat{\Pi}=\partial_{t}\hat{\Phi}$.
Eq.~\eqref{eq:bessel_ortog} then implies the usual commutation rules
\begin{equation}
	\label{eq:OP_com}
	\left[
		\hat{\Phi}(t,r),\hat{\Pi}(t,s)
		\right]
	=
	\frac{i\,\hbar}{4\,\pi\,r^{2}}\,\delta(r-s)
	\quad
	\iff
	\quad
	\left[
		\hat{a}_{k},\hat{a}_{p}^{\dagger}
		\right]
	=
	\frac{2\,\pi^{2}}{k^{2}}\,\delta(k-p)
	\ ,
\end{equation}
so that $\hat a_{k}$ and $\hat a_{k}^{\dagger}$ are annihilation and creation operators, respectively.
\par
The Fock space can be defined as usual starting from the vacuum $\hat a_{k} \ket{0}=0$,
which we associate with a Minkowski spacetime devoid of any physical events, as mentioned above.
Coherent states in this Fock space can be defined as eigenstates of the annihilation operators,
that is
\begin{equation}
	\label{eq:g_eingeq}
	\hat a_{k}\ket{g}
	=
	g_{k}\,e^{i\,\gamma(t)}
	\ket{g}
	\ .
\end{equation}
We can further set $\gamma(t)=k\,t$ to remove the time dependence,~\footnote{This approximation
	is valid for times $ \Delta t \sim k^{-1} $ and would break down in the presence of sufficiently high
	energy events.}
thus
\begin{equation}
	\label{eq:Phi_expct}
	\sqrt{\gn}\,\bra{g}\hat{\Phi}\ket{g}
	=
	\sqrt{\gn}\,
	\int_0^\infty
	\frac{k^{2}\,\d k}{2\,\pi^{2}}\,
	\sqrt{\frac{2\,\hbar}{k}}\, g_k\, j_0(k,r)
	=
	V_g(r)
	\ ,
\end{equation}
and the quantum versions of metrics of the form~\eqref{g} are given by
\begin{equation}
	\label{eq:g_function}
	-g_{tt}
	=
	g^{rr}
	=
	1+V_g(r)
	\ .
\end{equation}
The coherent state $\ket{g}$ therefore carries information about the spacetime
geometry encoded by the eigenvalues $g_{k}$ in Eq.~\eqref{eq:g_eingeq}.
\par
The metrics obtained from Eqs.~\eqref{eq:Phi_expct} and~\eqref{eq:g_function}
usually contain deviations from the classical solutions characterised by $V=V(r)$ that
they are built to reproduce.
This is because $\ket g$ must be normalisable in order to be well-defined~\cite{Casadio:2021eio}.
In particular, we can write
\begin{equation}
	\label{eq:gstate}
	\ket g
	=
	e^{-\frac{\mathcal N_g}{2}}
	\exp\left\{
	\displaystyle\int_{0}^{\infty}\frac{k^{2}\,\d k}{2\,\pi^{2}}\,
	g_k\,\hat{a}_k^{\dagger}
	\right\}
	\ket 0
	\ ,
\end{equation}
where the normalisation factor (or total occupation number) given by
\begin{equation}
	\label{eq:N}
	\mathcal{N}_g
	=
	\int_0^\infty
	\frac{k^{2}\,\d k}{2\,\pi^{2}}
	\left|g_{k}\right|^{2}
\end{equation}
must be finite.
However, one finds that classical black hole geometries correspond to coefficients
$g_k$ that make the above integral diverge both in the infrared (for $k\to 0$)
and the ultraviolet (for $k\to\infty$).
The former divergence is associated with the infinite volume of space, whereas the
latter appears because of the central classical singularity.
To avoid such divergences, the coefficients $g_k$ must necessarily depart from their classical
expressions in these two regimes and $V_g$ cannot be exactly equal to $V$.
\section{Quantum (A)dS}
\setcounter{equation}{0}
\label{Sect:QAdS}
The static dS patch~\eqref{gds} and the AdS spacetime~\eqref{gads} are characterised by
the metric function
\be
V_\Lambda
=
\frac{\Lambda}{3}\,r^2
\ .
\label{eq:Vr_cl}
\ee
We then proceed by expanding the function~\eqref{eq:Vr_cl} on spherical Bessel functions,
\be
\tilde V_\Lambda
&\!\!=\!\!&
\frac{4}{3}\,\pi\,\Lambda
\int_0^\infty
r^2\,\d r\,j_0(k\,r)\,r^2
\nonumber
\\
&\!\!=\!\!&
-\frac{2\,\pi\,i}{3\,k}\,\Lambda
\int_{-\infty}^{+\infty}
r^3\,\d r\,
e^{i\,k\,r}
\nonumber
\\
&\!\!=\!\!&
\frac{4\,\pi^2\,\Lambda}{3\,k}\,\delta^{(3)}(k)
\ ,
\label{tr2}
\ee
where $\delta^{(n)}$ denotes the $n^{\rm th}$ derivative of the Dirac delta distribution.
\par
The eigenvalues that determine the (A)dS coherent state $\ket{\Lambda}$ would be given by
\be
g_k
=
\sqrt{\frac{k}{2}}\,\frac{\tilde V_{\Lambda}(k)}{\lp}
\ ,
\ee
but $g_k^2\sim\left[\delta^{(3)}( k )\right]^2$ which determines the total
occupation number~\eqref{eq:N} for $\ket{\Lambda}$ is not well-defined even as a distribution.
The coherent state must therefore be regularised, for example, by replacing
\be
\delta(k)
\to
\frac{e^{-{k^2}/{\sigma^2}}}{\sqrt{\pi}\,\sigma}
\ ,
\ee
where $\sigma\sim 1/R_\infty>0$ acts as an IR cut-off scale associated with the (possibly)
infinite spatial volume.
This yields
\be
\tilde V_{\sigma\Lambda}
=
\frac{16\,\pi^{3/2}\,\Lambda}{3\,\sigma^5}
\left(3-2\,\frac{k^2}{\sigma^2}\right)
e^{-{k^2}/{\sigma^2}}
\ ,
\ee
and the regularised eigenvalues in the coherent state $\ket{\Lambda}$ correspondingly read
\be
g_{\sigma k}
=
\frac{16\,\sqrt{\pi^3\,k}\,\Lambda}{3\,\sqrt{2}\,\lp\,\sigma^5}
\left(3-2\,\frac{k^2}{\sigma^2}\right)
e^{-{k^2}/{\sigma^2}}
\ .
\ee
The total occupation number in the coherent state so defined is given by
\be
\label{eq:occN_integral}
\mathcal N_{\sigma\Lambda}
=
\int_0^\infty
\frac{k^2\,\d k}{2\,\pi^2}\,g_{\sigma k}^2
=
\frac{24\,\pi\,\Lambda^2}{9\,\lp^2\,\sigma^6}
\ .
\ee
Using $L^2=3/|\Lambda|$, we thus find that
\be
\mathcal N_{\sigma\Lambda}
\sim
\frac{R_\infty^6}{\lp^2\,L^4}
\label{occN}
\ee
diverges as the square of a spatial volume in units of $\lp\,L^2$.
\par
We can finally compute the quantum corrected metric function
\be
\label{eq:QR_func}
V_{\sigma\Lambda}
=
\int_0^\infty
\frac{k^2\,\d k}{2\,\pi^2}\,j_0(k\,r)\,\tilde V_{\sigma\Lambda}(k)
=
\frac{\Lambda}{3}\,r^2\,
e^{-\frac{\sigma^2\,r^2}{4}}
\ .
\ee
Note that we could formally take the limit $\sigma\to 0$ and recover the exact
classical expression~\eqref{eq:Vr_cl}, but there is no well-defined quantum state
$\ket{\Lambda}$ in such a limit.
\par
The (necessary) regularisation introduced above yields the quantum corrected metric
\begin{equation}\label{eq:QCmetric}
	\d s^{2}
	=
	-\left( 1-\frac{\Lambda}{3}\,r^{2}\, e^{-\frac{r^{2}\,\sigma^{2}}{4}}\right)
	\d t^{2}
	+
	\left( 1-\frac{\Lambda}{3}\,r^{2}\, e^{-\frac{r^{2}\,\sigma^{2}}{4}}\right)^{-1}
	\d r^{2}
	+
	r^{2}\,\d\Omega^{2}
	\ ,
\end{equation}
with significant consequences for the spacetime geometry.
The metric above does not solve the vacuum Einstein equations with $\Lambda$, as can be verified immediately by
plugging~\eqref{RsL} in~\eqref{eq:Einstein_eq}.
This is expected as our solution is constructed to reproduce that solution of the Einstein equations as close
as possible, while maintaining the well-definiteness of the quantum state~\eqref{eq:gstate}.
We can also compute the effective energy-momentum tensor $T_{\mu\nu}$ sourcing the quantum corrected
metric from the Einstein tensor $G_{\mu\nu}$ and obtain the effective energy density
\be
\rho
=
T^0_{\ 0}
=
-\frac{G^0_{\ 0}}{8\,\pi\,\gn}
=
\frac{\mpl\,\Lambda}{8\,\pi\,\lp}\,
e^{-\frac{r^{2}\,\sigma^{2}}{4}}
\left(
1-\frac{\sigma^2}{6}\,r^2
\right)
\ ,
\ee
the effective radial pressure
\be
p_r
=
T^1_{\ 1}
=
-\frac{G^1_{\ 1}}{8\,\pi\,\gn}
=
-\rho
\ ,
\ee
and the effective tangential pressure
\be
p_{\perp}
=
T^2_{\ 2}
=
-\frac{G^2_{\ 2}}{8\,\pi\,\gn}
=
-\frac{\mpl\,\Lambda}{8\,\pi\,\lp}\,
e^{-\frac{r^{2}\,\sigma^{2}}{4}}
\left(
1-\frac{\sigma^2}{12}\,r^2
\right)
\left(
1-\frac{\sigma^2}{2}\,r^2
\right)
\ .
\ee
Notice that when $ \sigma\to 0 \ \left( R_{\infty}\to\infty \right) $ the components reproduce exactly the
contribution of the cosmological constant term in the Einstein equations. The anisotropy ratio reads
\be
\Pi
=
\left|\frac{p_\perp-p_r}{p_r}\right|
=
\frac{\sigma^2\,r^2}{4}
\left|
\frac{10-\sigma^2\,r^2}{6-\sigma^2\,r^2}
\right|
\ ,
\ee
which vanishes for $r\to 0$ and reaches a value of $\Pi=9/20$ for $r=R_\infty=\sigma^{-1}$. In fact, the
metric~\eqref{eq:QCmetric} approaches asymptotically Minkowski at large $r$ (for $\sigma^2\neq 0$) faster than
any power, as can also be seen from the expression of the scalar curvature
\begin{equation}
	R_{\sigma\Lambda}
	=
	\frac{\Lambda}{12}\,e^{-\frac{r^{2}\,\sigma^{2}}{4}}
	\left(48-18\, r^{2}\,\sigma^{2} + r^{4}\,\sigma^{4} \right)
	\ ,
	\label{RsL}
\end{equation}
which vanishes exponentially for $r\to \infty $.
\begin{figure}[t]
	\begin{center}
		\includegraphics[width=0.45\textwidth]{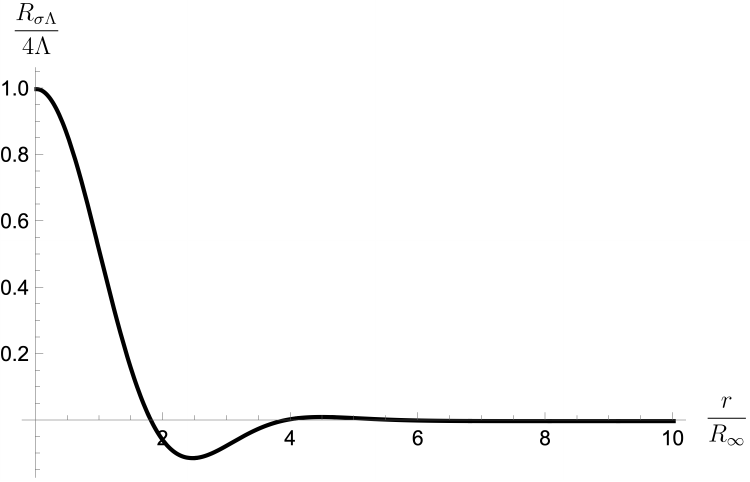}
		$\ $
		\includegraphics[width=0.45\textwidth]{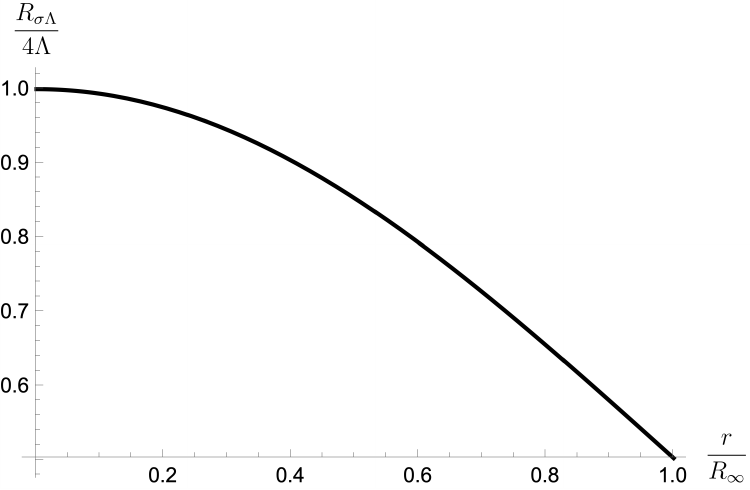}
	\end{center}
	\caption{Effective cosmological constant~\eqref{eqLeff}.}
	\label{Leff}
\end{figure}
\par
One can view the scalar~\eqref{RsL} as an effective cosmological constant, namely
\be
\Lambda_{\rm eff}
=
\frac{R_{\sigma\Lambda}}{4}
\ ,
\label{eqLeff}
\ee
which is plotted in Fig.~\ref{Leff}.
In particular, one can see that $\Lambda_{\rm eff}$ decreases towards $R_\infty=\sigma^{-1}$,
which violates the cosmological principle of homogeneity.
For dS, one must therefore assume that $R_\infty$ is large enough to satisfy the present bounds
on the scale of homogeneity of the visible Universe inside $L=\sqrt{3/\Lambda}$.
Provided that condition is satisfied, one obtains a local $\Lambda_{\rm eff}>\Lambda$
at cosmological scale (see right panel in Fig~\ref{Leff}).
In turn, this would imply that the local Hubble parameter is expected to be larger than the
cosmological one,
\be
\Lambda_{\rm eff}
=
H_{\rm local}
>
H_\Lambda
=
\Lambda
\ ,
\ee
which goes in the direction of the measured Hubble tension.
However, we remark that matter is totally neglected in the present analysis
as our goal here was to better understand the role of the cosmological constant alone.
In previous works, a black hole in dS background was considered, and the interplay between
matter and the cosmological constant was analysed in more detail.
The presence of a localised matter source leads to modifications in the ultraviolet and infrared behaviour
of the coherent state, and consequently Eqs.~\eqref{eq:occN_integral} and~\eqref{eq:QR_func}.
The possible connections of such modifications with the issues of Dark Matter and Dark Energy were analysed
in Ref.~\cite{Giusti:2021shf} with results similar to those found in Refs.~\cite{Cadoni:2017evg,Cadoni:2020izk}.
%
%
%
%
\section{Discussion}
\setcounter{equation}{0}
\label{Sect:concl}
In this work we have used the coherent state quantisation to construct a quantum state for maximally
symmetric spacetimes in four dimensions, that is dS and AdS.
In particular, we have shown that an infrared length scale $R_\infty$ is needed for the state to be
well-defined, and how this modification affects the spacetime geometry.
\par
We can further notice that setting $R_\infty = L$, the radius of curvature, results in the occupation
number~\eqref{occN} scaling as
\be
\mathcal{N}_{\sigma \Lambda}
\sim
\frac{L^{2}}{\lp^{2}}
\ ,
\ee
which is the same behaviour of the central charge that one finds by computing the 2-point function
of the stress tensor with the AdS/CFT correspondence, that is~\footnote{Odd dimensional CFT's do not have
	a central charge in the traditional sense, which is related to the trace anomaly of the stress
	tensor~\cite{deHaro:2000vlm}.
	However, in the holographic context,	it is customary to associate the central charge with the normalisation
	factor of the stress tensor 2-point function of the dual theory.}
\begin{equation}
	\label{eq:CFT_ct}
	C
	\sim
	\frac{L^{2}}{\lp^{2}}
	\ .
\end{equation}
A similar relation already appeared in the literature, for example in Ref.~\cite{Dvali:2013eja}.
Despite using different constructions for the quantum state, our results are compatible with the
AdS degrees of freedom being fundamentally related to those of a CFT in one dimension lower.
In fact, the expression of the central charge in Eq.~\eqref{eq:CFT_ct} is known exactly~\cite{Skenderis:2002wp},
from which one obtains
\begin{equation}
	R_{\infty}^{6}
	=
	\frac{2 L^{6}}{\pi^{3}}
	\ .
\end{equation}
One can also check that the occupation number for AdS in 5 dimensions displays the same behaviour as the
central charge.
It will be interesting to investigate if the quantum geometry produces other effects that
are known in the holography literature, or if this is just accidental.
\par
We should also notice that $ R_{\infty}=L $ is ruled out for the dS spacetime in order to
match current observations, but there is no such restriction in the case of AdS.
Whether this is an indication that this holographic connection is exclusive to AdS, or just a
coincidence remains an open question.
\par
Finally, we remark that a setup including matter in the cosmological context, that is coherent states for
the Friedmann-Lema\^{i}tre-Robertson-Walker models, is technically possible, but requires coherent states
with time dependence, which is currently under construction and are left for future research.
\subsection*{Acknowledgements}
P.M.~thanks FAPESP for the financial support (grants No. 2022/12401-9 and No. 2023/12826-2).
A.G.~is partly supported by the Science and Technology Facilities Council
(grants n.~ST/T006048/1 and ST/Y004418/1).
A.G.~and R.C.~carried out this work in the framework of the activities of the
Italian National Group of Mathematical Physics [Gruppo Nazionale per
		la Fisica Matematica (GNFM), Istituto Nazionale di Alta Matematica (INdAM)].
R.C.~is partially supported by the INFN grant FLAG.
\newpage
\bibliographystyle{unsrt}
\end{document}